\documentclass[11pt]{article}

\usepackage[utf8]{inputenc}
\usepackage[T1]{fontenc}
\usepackage[letterpaper,margin=1in]{geometry}
\usepackage{amsmath,amssymb}
\usepackage{graphicx}
\usepackage{placeins}   % provides \FloatBarrier
\usepackage{microtype}
\usepackage{fancyhdr}
\usepackage{caption}
\usepackage{hyperref}
\hypersetup{colorlinks=true,linkcolor=black,citecolor=black,urlcolor=black}

\usepackage{titlesec}
\titlespacing*{\section}
  {0pt}
  {2.0ex plus 0.6ex minus 0.2ex}
  {1.0ex plus 0.2ex}

\newcommand{\Mw}{M_{\mathrm{weak}}}

\newcommand{\sw}{\sin^{2}\!\theta_{W}}

\title{\bfseries
The BSM Rollercoaster, 1980--2000\thanks{Contribution to the 4th International Symposium on the History of Particle Physics, CERN, November 2025.}\\[0.4em]
\large A Participant's Perspective}

\author{Savas Dimopoulos \\[2pt]
\textit{\normalsize Leinweber Institute for Theoretical Physics, Stanford University}}
\date{}

\begin{document}
\maketitle
\thispagestyle{fancy}

\smallskip
\noindent{\small
\textbf{Abstract.}
This contribution is a personal historical retrospective on
beyond-the-Standard-Model (BSM) physics from 1978 to the early 2000s,
organized around its emotional rollercoaster. I trace two waves
of optimism separated by a multiyear contraction of confidence, both ultimately ending in disappointment when decisive experimental confirmation failed
to arrive.

The story begins in 1978, when the Standard Model emerged as the front-runner. It then moves from the first wave of excitement over naturalness and Technicolor, rumored proton-decay events, and the construction of the Minimal Supersymmetric Standard Model, to the contraction of confidence in the late 1980s. A dramatic revival came in 1991, when the LEP and SLC precision data pointed toward supersymmetric unification, but it eventually gave way to renewed disappointment over the absence of superpartners and the sharpening of the LEP paradox.

I also describe how the realization that Newtonian gravity had been
tested only down to distances of order centimeters opened the door to
large extra dimensions. Finally, I discuss how Weinberg's 1987 anthropic
argument for the cosmological constant, the 1998 discovery of cosmic
acceleration, and the emergence of the string landscape changed the
intellectual status of naturalness itself and led to
Split Supersymmetry.

I have tried to describe not only what was proposed, but why these ideas
felt compelling at the time, why confidence rose and fell, and how later
evidence changed our sense of what could count as an explanation.

\par}
\medskip

\section{Introduction}

I want to describe the emotional rollercoaster of
beyond-the-Standard-Model physics during the 1980s and 1990s.  Both
decades began with great excitement and both ended with disappointment,
each time because the hoped-for experimental confirmation never came.  That was, in a real sense, the story of our lives during those
twenty years.

This article grew out of a symposium on the History of Particle Physics
in the period 1980--2000, the fourth in a series and the first in more than thirty years. The organizers gave the speakers an unusual
and welcome charge.  We were explicitly asked \textit{not} to give just a
physics talk, of the kind we might have given at the time; the published
literature already preserves those.  We were asked instead to convey what
the literature often does not: why certain ideas gained ground, how
choices were made under uncertainty, and how mistakes, accidents, dead
ends, and emotional turns shaped the period.  The audience was unusually
broad and alive to this kind of question: senior physicists, historians
and philosophers of science, science reporters, current and former
laboratory directors, and even politicians.

I have tried to honor that charge by concentrating on episodes I knew
from the inside---my own work, the work of close colleagues, and the
undercurrents that rarely make it into print.  The resulting account is
necessarily selective.  Its purpose is not to give a comprehensive
technical review, but to contribute a participant's testimony about how
the field looked while its future was still unknown.  Although the main
focus is the period 1980--2000, the story naturally spills a little into
the early 2000s, because some of the consequences of that period only
became clear then.

\begin{figure}[htbp]
\centering
\makebox[\textwidth][c]{%
  \includegraphics[width=1.15\textwidth]{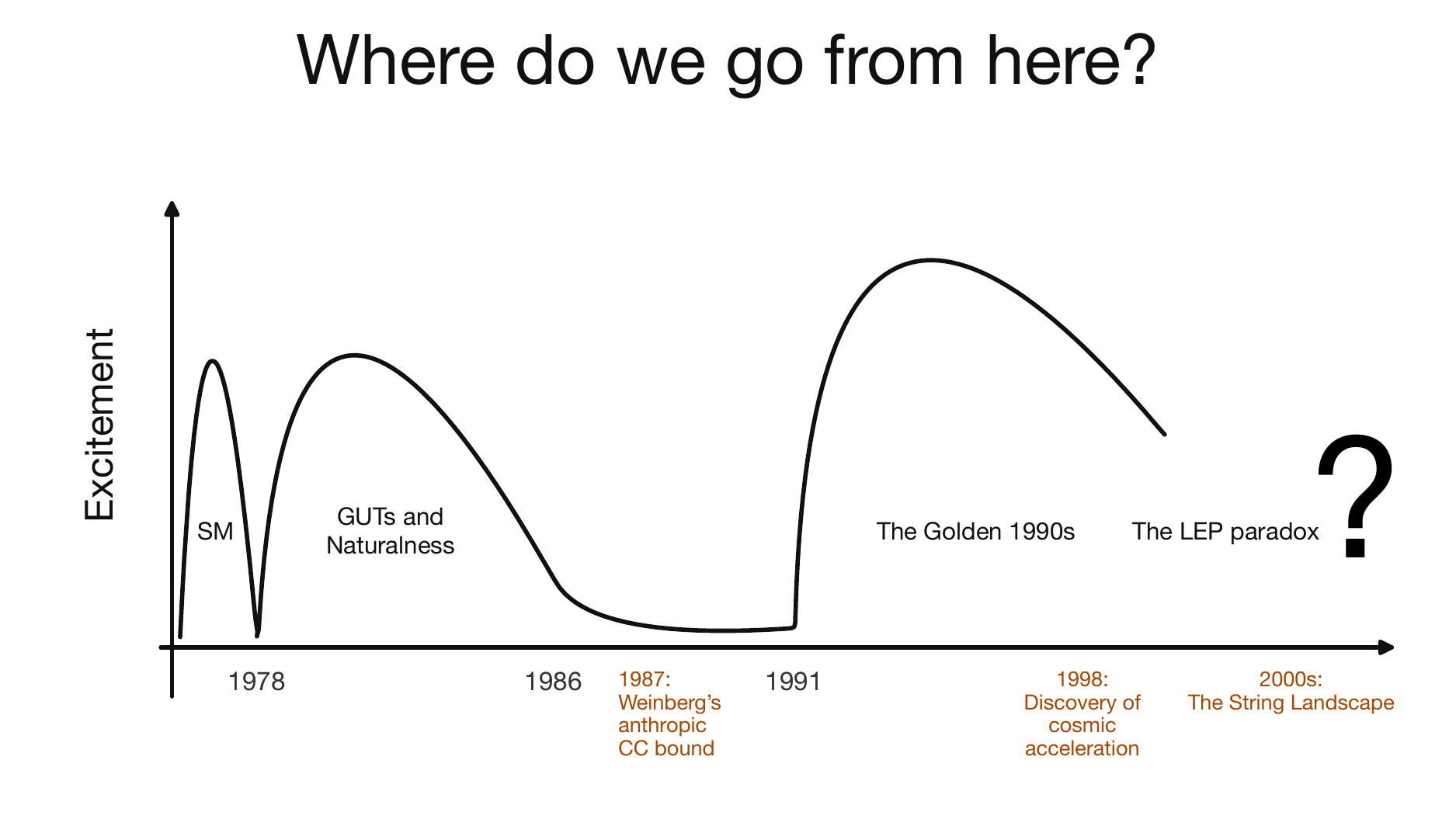}%
}

\caption{The emotional rollercoaster of the naturalness and unification
program in beyond-the-Standard-Model physics from the late 1970s to the
early 2000s.  The vertical axis is
``excitement,'' or perceived promise, not a measured quantity.  The curve
marks the two main waves discussed in the text, while the colored labels
below the axis mark the cosmological-constant and landscape story.  The
title ``Where do we go from here?'' is meant historically.}

\label{fig:rollercoaster}
\end{figure}

The most compact way to state the story is Figure~\ref{fig:rollercoaster}.
It is an intentionally subjective plot.  
The curve chiefly reflects the naturalness and unification program in which I worked; other areas, notably neutrino physics, followed different trajectories.
Large Extra Dimensions, which arrived in 1998, are not marked on it; they opened a different track rather than reversing this one. 
The horizontal axis is time; the
vertical axis is not a measured quantity, but the perceived promise of a
coherent path beyond the Standard Model, as experienced by many theorists of my generation.  
The rollercoaster was conceptual as well as emotional: the preferred form of explanation moved from symmetry and dynamics, through geometry, to the more controversial idea of environmental selection.
The peaks were produced not only by attractive theories,
but by moments when theory and experiment seemed to be converging; the
valleys came not only from negative results, but from the loss of that
convergence.  The rest of this article follows this curve.  It is the
spine of the story.  The units are subjective, but many theorists of my
generation will recognize the curve.

\section{1978: The Standard Model becomes the front-runner}

The story begins in mid-1978, when the SLAC electron--deuteron
scattering experiment \cite{prescott} observed parity violation in
neutral currents.  The measured asymmetry agreed strikingly well with the
$SU(3)\times SU(2)_L\times U(1)$ gauge theory that we now call the
Standard Model.  The previous several years had been a fertile period
for model building because the data still looked messy: the high-$y$
anomaly in neutrino scattering, the apparent absence of atomic parity violation, and continuing controversies over the
interpretation of neutral-current experiments.  My generation was trained,
very effectively, to build models in such an environment.  I was a
graduate student at the University of Chicago at the time.

When the Standard Model became the front-runner---``established'' would
be too strong a word for those first weeks---the students and postdocs
around me were elated for a few days.  Then we became depressed.  The
natural question was: what do we do now?  
Some of my contemporaries from Chicago left physics, and I
almost did so myself.

This was not just a manner of speaking. 
I had two postdoctoral offers: three years at Berkeley and two at Columbia. I chose the shorter Columbia appointment, partly because New York put Wall Street close at hand if physics did not work out.
As it happens, 1978 was not a
bad year for that plan: it was the year Jim Simons started what became
his extraordinary investment career. Simons, of course, later
created Renaissance Technologies and the Simons Foundation.

Within a few months the despair began to dissipate as the central
questions shifted abruptly.  Until 1978 the question had been \textit{what} is the
Standard Model?  Now it became \textit{why} is the Standard Model the way it
is?  Why is the cosmological constant so small?  Why are the fermion
masses what they are?  Why is there more matter than antimatter?  Of all
these questions, the most fruitful for particle theory turned out to
be the hierarchy problem: why is gravity so weak?  That question drove
the construction of beyond-the-Standard-Model theories more strongly than
any other.

\section{Naturalness and the hierarchy problem}

The hierarchy problem can be illustrated by the simplest experiment in
the world.  When I lift my arm, the electrical forces in my muscles
overcome the gravitational pull of the entire Earth.  In particle
language, the same puzzle is expressed by saying that the weak scale is
seventeen orders of magnitude smaller than the Planck scale.  Why should
the scale of the weak interactions be so tiny compared with the scale at
which gravity becomes strong?

This is the classical version of the problem.  The quantum version is
both subtler and more powerful.  The Higgs is a
scalar, and scalar masses are not automatically protected against the
influence of ultraviolet physics.  If the Standard Model is embedded in a
larger theory containing heavy particles---a grand unified theory,
string theory, or almost any larger framework one might have imagined in the late twentieth century---then the Higgs mass is sensitive to those
heavy thresholds.  The weak scale is therefore not insulated from the
ultraviolet.  Without a protective mechanism, keeping the Higgs light
requires an enormous cancellation between apparently unrelated
contributions.  If the relevant high scale is the Planck scale, the
cancellation has to be fantastically precise, at the level of roughly one
part in ten to the thirty-four.

The name we used for the desired cure was \emph{naturalness}.  Small
numbers in physics, especially very small dimensionless ratios, should
not simply be inserted by hand.  They should arise from a symmetry, a
dynamical mechanism, or some other structural reason.  Naturalness
therefore transformed a collection of puzzles into a research program.
It was not merely an aesthetic preference.  It became the guiding logic
behind much of beyond-the-Standard-Model physics for the next few
decades.

It is worth realizing that this logic was not obvious from the beginning. The radiative version of the hierarchy problem was subtle enough that even Steven Weinberg took seriously a possible escape. 
Around 1979, as I recall, Weinberg privately circulated a preprint to a small group of friends, arguing that the hierarchy problem could be removed by dimensional regularization.

The point, however, is not the regulator.  
Dimensional
regularization can hide explicit power divergences, but it cannot remove
the physical effect of actual heavy particles.  Once the Standard Model
is enlarged to include such particles, the Higgs still feels them.
Weinberg recognized the problem, withdrew the preprint, and asked his
friends to tear up their copies.  I include this episode not as a
curiosity, but because it shows how ideas that now seem elementary were
still being clarified in real time.  The hierarchy problem became
compelling only after the community learned to distinguish a formal cutoff
artifact from a genuine sensitivity to heavy physical thresholds.  Even the cleanest ideas were not always clean at birth.

\section{Five approaches to naturalness}

Five responses shaped the next quarter-century.

\smallskip\noindent
\textbf{Technicolor} (1978).  The elementary Higgs is replaced by a
condensate of new strongly interacting fermions \cite{technicolor}.
The weak scale is then generated dynamically, much as the QCD scale is
generated from a weak high-energy coupling.

\smallskip\noindent
\textbf{Softly Broken Supersymmetry} (1981).  The Higgs is
protected from short-distance physics by supersymmetry, which pairs it
with a fermionic partner, the Higgsino.

\smallskip\noindent
\textbf{Large Extra Dimensions} (1998).  The four-dimensional Planck
mass is not fundamental; the true quantum-gravity scale is of order a
TeV.  The apparent weakness of gravity is geometrical, arising from its
dilution into large compact extra dimensions.

\smallskip\noindent
\textbf{Warped Extra Dimensions} (1999).  A warped extra dimension
generates an exponential hierarchy between the Planck and weak scales
\cite{rs}.  In the holographic language, this is closely related to
strongly coupled, nearly conformal versions of technicolor
\cite{Holdom:1981rm}.

\smallskip\noindent
\textbf{Split Supersymmetry} (2004).  Naturalness for the Higgs mass is
given up.  The weak scale is selected anthropically in a landscape of
vacua, while gauge coupling unification and a dark-matter candidate are retained.

This list is useful because it shows both the power and the evolution of
the naturalness idea. I will not try to review all five.  I will concentrate on the three in
which I was personally involved in a constructive way: Softly Broken
Supersymmetry, Large Extra Dimensions, and Split Supersymmetry.  I have not worked on warped compactification, the influential idea of Randall and Sundrum, so I will not attempt to do it justice here.  I should nevertheless say a word about technicolor,
because its impact was enormous.  Through Weinberg's early dynamical
symmetry-breaking paper and especially Leonard Susskind's landmark
technicolor paper \cite{technicolor}, it essentially started the field of
TeV BSM model building.  It taught us to think of the weak scale as a
dynamically generated scale, not merely as an input parameter.

The trouble came when technicolor was extended to give masses to ordinary
quarks and leptons.  Extended Technicolor \cite{etc} was
introduced for exactly this purpose, but it generically produced dangerous
flavor-changing neutral currents \cite{dimopoulosellis}.  My own
contributions to technicolor were mostly useful in this negative sense.
They helped clarify how hard the flavor problem was, and they also 
redirected me toward supersymmetry.

\section{The path to the Minimal Supersymmetric Standard Model}

After the disappointment with technicolor, I began learning
supersymmetry.  It had been discovered in the early 1970s, and its
central virtue for the hierarchy problem was already clear.  Fermion
masses can be protected by chiral symmetries.  Supersymmetry extends
this protection to scalars: the Higgs comes with a fermionic partner,
the Higgsino, and in the supersymmetric limit their masses are tied
together.  A light Higgs would then inherit the protection enjoyed by a
light fermion.  It would not be a miracle.

The historical question is then: why had no phenomenologically viable
supersymmetric Standard Model emerged before 1981?  
Pierre Fayet had constructed supersymmetric extensions in the 1970s,
recognized the need for two Higgs doublets, and, with Glennys Farrar,
introduced R-parity \cite{fayet}, but none of these models had a
phenomenologically acceptable spectrum.
A central obstacle was the supertrace theorem of Ferrara,
Girardello, and Palumbo \cite{supertrace}.  In the simple tree-level
constructions then under consideration, the theorem forced an
unacceptable spectrum or an unacceptable vacuum.  In particular,
charged or colored scalars could acquire vacuum expectation values at
the weak scale, giving the photon or the gluons masses of order $\Mw$.
That was fatal.

The key to evading this obstruction came from the hierarchy problem
itself.  It had taught us that the tree-level spectrum need not be the
last word.  The Higgs was the extreme example.  Its bare mass is not a
physical observable; what matters is the mass after quantum corrections,
and for a scalar those corrections can be so large, because of its
sensitivity to very heavy physics, that they dominate the tree-level
value.  That is the essence of the hierarchy problem.  But the same
general lesson---that quantum effects can dominate tree-level scalar
masses---suggested a way around the supertrace theorem, which constrains
only the tree-level spectrum.  Why not then build supersymmetric theories
in which the scalar masses constrained by the theorem are likewise
dominated by quantum effects associated with supersymmetry breaking?  In
such theories the unacceptable tree-level spectrum would not be the
physical spectrum.  It would be replaced by one with radiatively
generated soft masses.

This line of thought led to three papers in the spring of 1981.

\smallskip\noindent
\textbf{Supercolor}, with Stuart Raby \cite{supercolor}, was an attempt
to break supersymmetry dynamically through a new strong interaction
analogous to technicolor, and thereby to generate a viable scalar
spectrum.  It was an early precursor of later work on dynamical
supersymmetry breaking and gauge mediation.  Closely related ideas were
also pursued by Witten, and by Dine, Fischler, and Srednicki
\cite{wittendsb,Dine:1981za}.

\smallskip\noindent
\textbf{Supersymmetry and the scale of unification}, with Stuart Raby
and Frank Wilczek \cite{drw}, set aside the mechanism of supersymmetry
breaking and asked a simpler question: what happens to grand unification
if the minimal supersymmetric particle content appears near the weak
scale?  The running of the couplings is then qualitatively different
from the non-supersymmetric case.  Related analyses and constructions by
 Ib\'a\~nez and Ross, by Einhorn and Jones, and by Sakai became part of the
same emerging picture \cite{Ibanez:1981yh}.

\smallskip\noindent
\textbf{Softly broken supersymmetry and SU(5)}, with Howard Georgi
\cite{dg}, took the opposite approach.  Instead of deriving
supersymmetry breaking, we inserted soft supersymmetry-breaking terms by
hand and studied the resulting low-energy theory.  This separation
between the unknown mechanism of supersymmetry breaking and the
low-energy soft terms was extremely useful: it turned supersymmetry into
a concrete phenomenological framework that experimentalists could
search for, and it helped define the collider searches that followed.

\medskip
Together these papers helped bring what became the Minimal Supersymmetric
Standard Model (MSSM) into a phenomenologically viable form and
established much of its key phenomenology.  For future experiments, the
framework predicted a roughly doubled particle spectrum.  For
experiments already running in 1981, it had two immediate consequences.
First, the unification scale was roughly a factor of thirty higher than
in non-supersymmetric $SU(5)$, suppressing the conventional
dimension-six proton-decay rate beyond the reach of the experiments then
under way.  Second, the predicted weak mixing angle $\sw$ was shifted
upward by about ten percent relative to the non-supersymmetric grand unified
prediction.  Those two predictions---no immediate proton decay from
gauge-boson exchange and a different weak mixing angle---were how the
new theory first faced experiment.

It is hard, in retrospect, to convey how concentrated those months were.
The three papers were written in close succession, with overlapping
collaborators, each addressing a different face of the same problem:
the dynamical breaking of supersymmetry, the running of couplings, and
the low-energy soft spectrum.  At least in my mind, they were not
separate projects but pieces of one structure that was only beginning to
take shape.  The MSSM did not emerge from a single paper; it crystallized
across several papers, conversations, and calculations.

\section{1981--1989: from the Second to the Last Workshop on Grand Unification}

The workshops on grand unification were the high points of the BSM
calendar in the early 1980s.  I went, as a young faculty member, to the
Second Workshop on Grand Unification in Ann Arbor, Michigan, on April
24--26, 1981 \cite{annArbor81}, where I had been invited to present the new supersymmetric work
alongside Howard Georgi.

\begin{figure}[htbp]
\centering
\makebox[\textwidth][c]{%
  \includegraphics[width=1.15\textwidth]{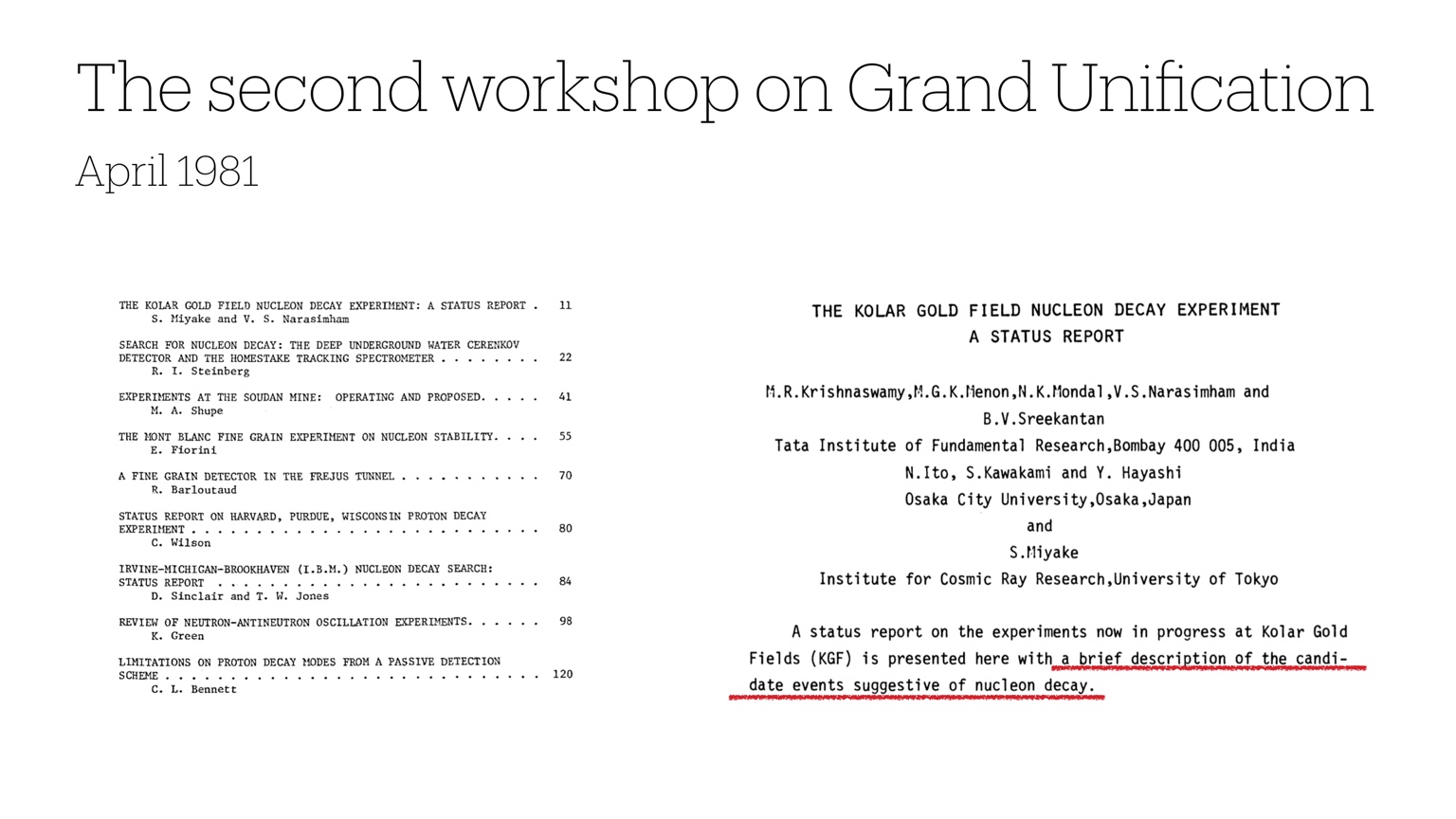}%
}
\caption{The Second Workshop on Grand Unification, Ann Arbor, April
24--26, 1981 \cite{annArbor81}.  Left: a proceedings page listing nucleon-decay status reports.
Right: the opening of the Kolar Gold Field status report, describing
``candidate events suggestive of nucleon decay''; emphasis added.
None of these events was confirmed as nucleon decay.}
\label{fig:workshop}
\end{figure}

On the first day I learned, with some discomfort, that there were rumors
of candidate events in some proton-decay experiments.
Figure~\ref{fig:workshop} reproduces the proceedings contents page and
the opening of the Kolar Gold Field status report, which presented three
candidate events described as ``suggestive of nucleon decay''
\cite{annArbor81}.  This was not ideal news for a young
theorist about to argue that the experiments then running should not see
proton decay.  Briefly, I worried about my career.

I gave the talk anyway, and nervously.  The other half of the prediction
concerned the relation between $\sw$ and the strong coupling
$\alpha_3$.  At the time, the central value favored the
non-supersymmetric prediction, though with large error bars; the
supersymmetric prediction sat near the edge.  The combination of no
observable proton decay and an apparently wrong weak mixing angle was
a difficult start, for supersymmetry and for me.  After my talk, the most
encouraging words I remember from a senior person came from Steven
Weinberg, in his conference summary \cite{annArborSW}.  He liked the theory anyway.

Ironically, our prediction of no observable proton decay itself rested on an
incomplete argument.  We had considered only the usual
gauge-boson-mediated dimension-six operators.  We missed a point that
became obvious in retrospect: in supersymmetric grand unified theories
there are also dimension-five operators, mediated by heavy color-triplet
Higgsinos, which can readily produce proton-decay rates within
experimental reach.  This was pointed out soon afterward by Weinberg,
and by Sakai and Yanagida \cite{dim5}.  The rumored proton-decay signals,
meanwhile, soon evaporated.

After Ann Arbor, activity in supersymmetry grew very rapidly.  There
were many important developments.  One of the most important was
supergravity mediation, an elegant mechanism by which supersymmetry
breaking in a hidden sector could be communicated to the visible sector
through interactions of gravitational strength \cite{sugra}.  

I also want to single out an important 1986 paper by Hall, Kosteleck\'y, and Raby \cite{hkr}. It pointed out an early dark cloud in the supersymmetry (SUSY) story. It showed that interactions introduced to explain the observed hierarchy of quark masses could generically spoil squark degeneracy or alignment, thereby inducing dangerous low-energy flavor-changing neutral currents. The problem that had severely damaged technicolor returned, in milder but still troubling form. This paper had a significant impact on Howard Georgi; it was one reason he moved away from supersymmetry and back toward technicolor.

But the main reason for the fading enthusiasm of the late 1980s was
simpler: neither of the two linked ideas, grand unification and
supersymmetry, had shown up experimentally.  There was no proton decay,
and there were no superpartners.

So by the latter part of the 1980s the optimism of the early decade had
faded, even as the superstring revolution was generating a different
surge of excitement elsewhere.  The annual Workshops on Grand Unification themselves came to an
end.  The 1989 meeting was titled, with characteristic bluntness,
\emph{The Last Workshop on Grand Unification}.  In the volume's
epilogue \cite{frampton}, Paul Frampton wrote: ``Alas, none of the
principal predictions of grand unified theories have been confirmed.''
That was written in August 1989---the same month LEP recorded its first collisions,
and four months after SLC recorded its first $Z$.  History has a
sense of humor.

\section{LEP/SLC: Supersymmetry in the loops?}
\label{sec:lep}

\begin{figure}[!t]
\centering
\makebox[\textwidth][c]{%
  \includegraphics[width=1.15\textwidth]{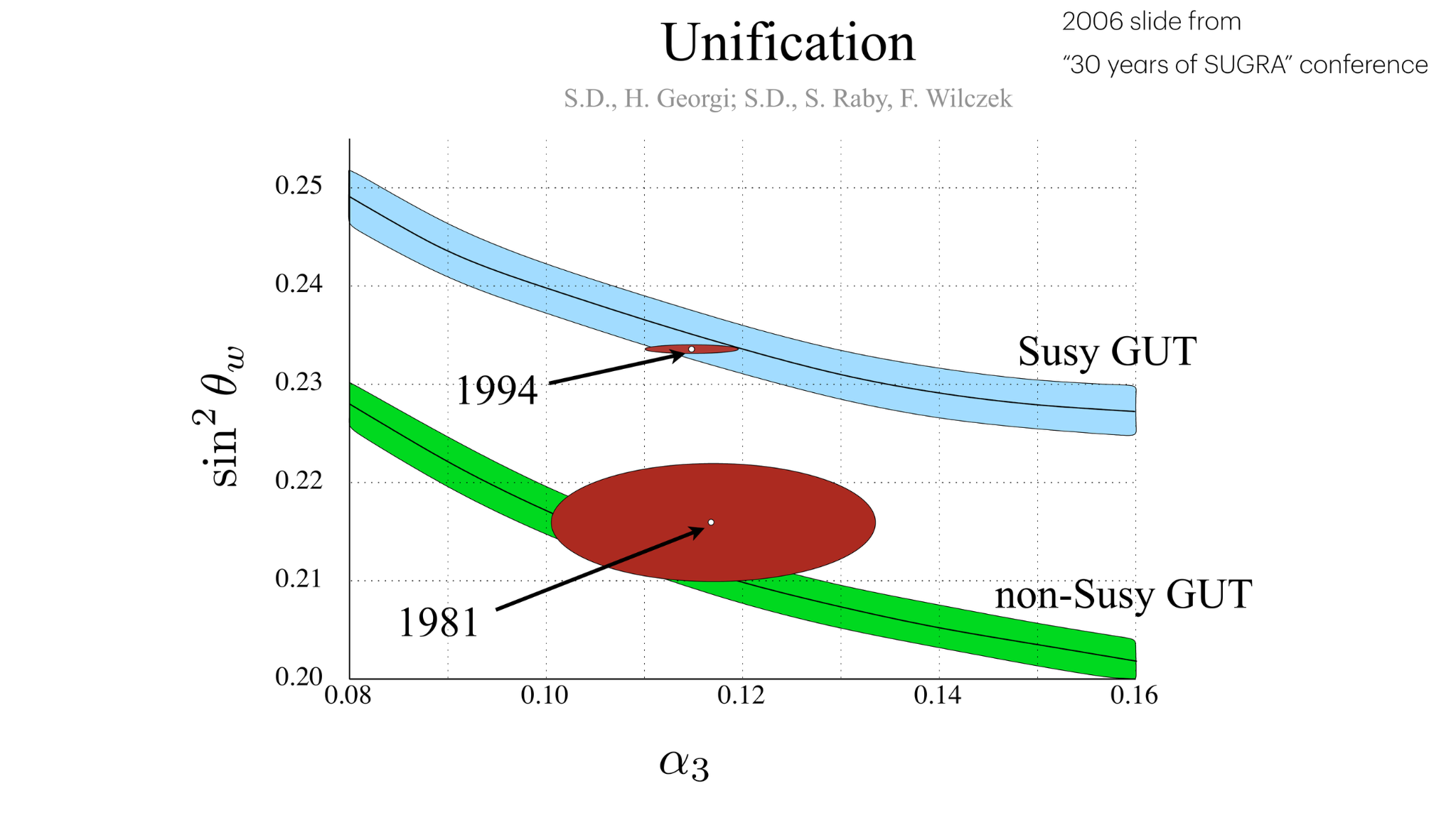}%
}
\caption{Gauge coupling unification in the $\sin^{2}\!\theta_W$--$\alpha_3$
plane.  The blue and green bands are the supersymmetric and
non-supersymmetric $SU(5)$ predictions \cite{drw,dg}.  The large 1981
ellipse favored the latter, though the errors were large; the small 1994
ellipse lay on the supersymmetric band.  The empirical ellipses are
reproduced from my 2006 retrospective slide for the ``30 Years
of SUGRA'' conference; for the 1991 LEP/SLC analyses, see
Ref.~\cite{amaldi}.}
\label{fig:unification}
\end{figure}

The precision electroweak data from LEP and SLC changed the mood almost
overnight.  In 1991 several papers took the new precision measurements, together with the measured strong coupling, and ran the gauge
couplings up to high energies \cite{amaldi}.  They found that the
electroweak and strong couplings, extrapolated by the renormalization
group, failed to unify in the non-supersymmetric Standard Model but came
together strikingly well in the MSSM\@.  Figure~\ref{fig:unification}
shows the transformation.  The large 1981 ellipse was replaced by a
small ellipse that had moved directly onto the supersymmetric band.

This produced enormous excitement.  The natural reading, at least for
many of us, was that we had already seen virtual supersymmetric
particles---not in the detector, but in the loops.  The superpartners
were contributing to quantum corrections and altering the running of
the couplings.  Surely, we thought, if the virtual particles were
already visible, the real ones could not be far away.

It is worth emphasizing why this felt so persuasive.  The three gauge
couplings, precisely determined at low energies, were extrapolated
upward by roughly fourteen orders of magnitude in energy.  In the
non-supersymmetric Standard Model the couplings nearly meet, but miss
by much more than the experimental uncertainties.  In the MSSM, once
the superpartners enter near the weak scale, the three couplings meet
near $2\times 10^{16}$~GeV to within a few percent. 
The agreement was not unique, and threshold corrections at both the weak and GUT scales could shift it; within the minimal spectrum, however, it was remarkably economical.
This did not feel
like numerology.  It connected the weak scale to a grand unified scale
through quantum corrections, and it did so with just the particle
content required to protect the Higgs mass.  The temptation to read
this as confirmation rather than suggestion was very strong.  Many of
us yielded to that temptation.

By the late 1990s, however, no superpartners had been seen directly at
LEP, and discomfort began to set in.  With time this became part of
what is now often called the LEP paradox: naturalness, reinforced by
gauge coupling unification, pointed strongly toward new physics near
the weak scale, while the precision observables remained stubbornly
Standard-Model-like and direct signatures refused to appear
\cite{Barbieri:2000gf}.  That tension became one of the central
emotional facts of the field.

At this point the chronology branches. While the LEP paradox was sharpening, two ideas were maturing into alternative approaches: short-distance gravity and anthropic cosmology. I take them in turn before returning to supersymmetry and showing how anthropic reasoning reshaped it in Split Supersymmetry.

\section{Large Extra Dimensions}

I now step back to 1988 to trace three developments that led to Large Extra Dimensions. It is a good illustration of how synthetic the scientific process is.

The first development came in July 1988, when Ignatios Antoniadis and I were attending a conference in Crete. We were swimming under the full moon when we first asked: what if the fundamental string scale is a TeV rather than the Planck scale?
We continued thinking about this during our summer meetings in Greece in 1988, 1989, and 1990, but did not get too far. Ignatios was brave enough to write a 1990 paper on a closely related precursor---a compactification dimension opening at a few TeV, though not yet a TeV fundamental string scale \cite{antoniadis90}. I felt that the ideas were not yet adequately developed and did not participate, despite his insistence. In retrospect that may not have been the right call.

\begin{figure}[htbp]
\centering
\makebox[\textwidth][c]{%
  \includegraphics[width=1.15\textwidth]{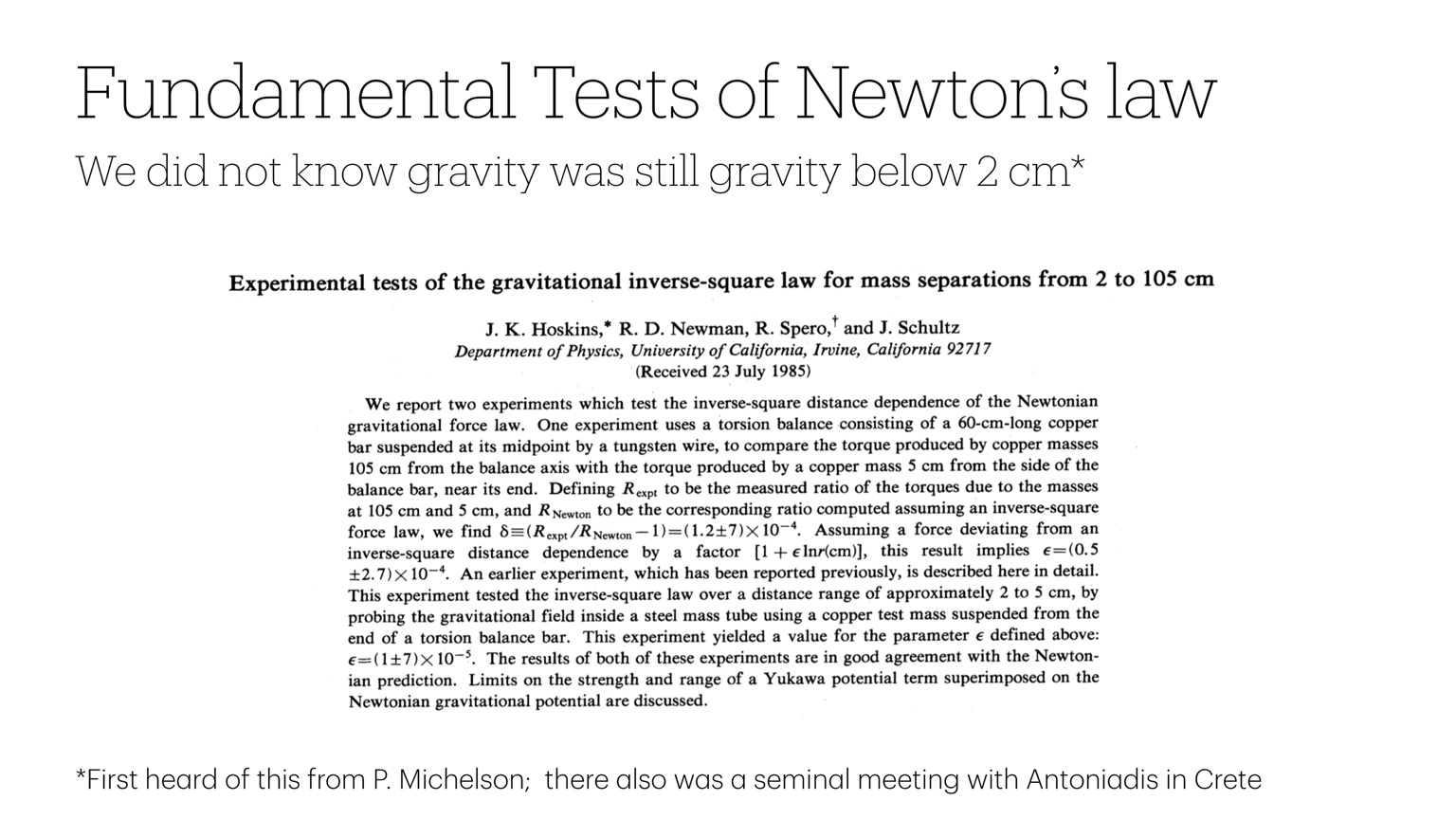}%
}
\caption{A slide built around the published abstract of Hoskins, Newman,
Spero, and Schultz (1985) \cite{hoskins}, the most precise test of the
Newtonian inverse-square law available at the end of the 1980s.  The
smallest mass separation tested was about two centimeters.  Below that
distance, laboratory tests of gravity were largely unexplored.  This was
the empirical opening that Peter Michelson pointed out to me, and it
became one of the seeds of large extra dimensions.}
\label{fig:hoskins}
\end{figure}

The second development came from my Stanford colleague Peter Michelson, who
pointed out to me a startling fact.  The best laboratory tests then available---particularly the measurements of Hoskins, Newman, Spero, and Schultz shown in Figure~\ref{fig:hoskins} \cite{hoskins}---had tested Newton's inverse-square law only down to separations of about two centimeters. Below that distance, laboratory gravity was largely
unexplored.  This struck me as an astonishing window of opportunity: a
huge untested range in which new, gravitational-strength physics could
in principle have been hiding.

I could not get this thought out of my mind and would mention it to every physicist I met. It opened up a universe of possibilities for new physics that I began exploring. In the mid-nineties I was trying to motivate my experimental colleagues, including Aharon Kapitulnik, to search this regime. In separate papers with Gian Giudice, and with Ignatios Antoniadis and Gia Dvali, I explored how light scalar fields, including moduli, could mediate new forces of gravitational strength at sub-centimeter distances \cite{moduliforces}.

The third significant development happened with Gia Dvali at the CERN cafeteria in December of 1996. Gia had just posted a paper with Misha Shifman \cite{dvalishifman} and I wanted to explore it. In the course of that conversation we hit on a key idea: perhaps gravity can spread into the space of extra sub-centimeter dimensions, while all other forces and familiar particles remain confined to our three dimensions. We decided to work on putting these three ingredients---TeV, cm, and the spreading of gravity---together, after finishing some other projects. I arranged not to teach in the winter of 1998, so that I could go to Paris to learn more about extra dimensions from Ignatios. I also invited Ignatios and Gia to visit me at Stanford later that winter, so that the three of us could work on it together.

In the fall of 1997, Nima Arkani-Hamed came to SLAC as a postdoc. We immediately hit it off and had great fun exploring wild ideas. We talked about the extra-dimensional ideas, but concentrated mostly on the cosmological constant problem---whether somehow we could dilute its strength by spreading it or ``off-loading'' it into extra dimensions. I invited him to join me in Paris that winter to learn more about extra dimensions from Ignatios. It took Nima about one whole second to agree to come to Paris, for the sake of science.

We had an incredibly productive time in Paris, though it was not obvious in the moment---there were too many exciting new possibilities, and it took time to disentangle them. The core idea had not yet crystallized. When we returned to Stanford and Gia joined us, it rose to the surface \cite{add1}. For the next six months I had one of the most thrilling stretches of my career. We all knew we were working on something remarkable, and we had it entirely to ourselves---like kids let loose in a toy store. We explored all kinds of possibilities, science fiction on solid mathematical footing. Ignatios visited Stanford for a few weeks and showed us how to embed these ideas in string theory on even firmer ground \cite{add2}. After this, Nima and I at Stanford, and Gia from a distance, worked incredibly hard for more than three months straight to make sure that these models were not excluded by cosmological, astrophysical, or laboratory data \cite{add3}.

My first seminar on this outside Stanford was at the Institute for Theoretical Physics in Santa Barbara in June of 1998. There was a string workshop there, and many top string theorists were present. The reception was enthusiastic. After my second talk at the SUSY 98 conference in Oxford that July, interest in these ideas exploded and lasted for many years. The work inspired new experiments---small-scale searches for new sub-millimeter dimensions and forces, and collider searches for new particles and energy leaking into extra dimensions.

\begin{figure}[!tbp]
\centering
\includegraphics[width=0.90\linewidth]{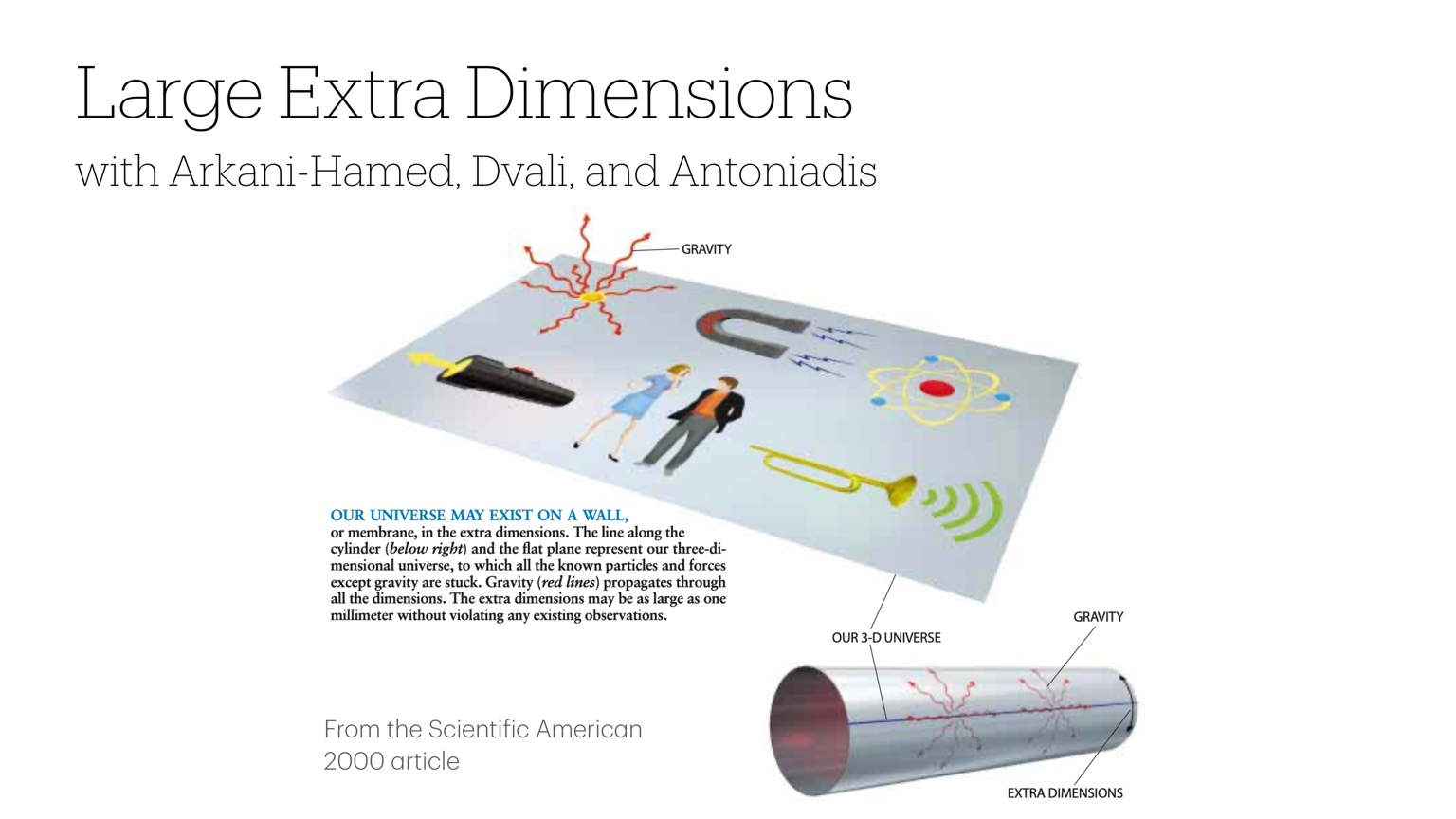}
\caption{The Large Extra Dimensions picture, in the cartoon form that
helped make the idea immediately understandable.  The plane, and
equivalently the line on the cylinder, represent our $3+1$ dimensional
world, to which the known particles and non-gravitational forces are
confined.  Gravity, by contrast, propagates in all dimensions.  Its
apparent weakness is then a geometrical effect: its flux is diluted into
the extra-dimensional volume.  The image is adapted from the 2000
\emph{Scientific American} article \cite{sciam}.}
\label{fig:led}
\end{figure}

The fact that the inverse-square law had been tested only down to the
centimeter scale meant that the extra dimensions could be enormously
larger than the Planck length, as large as a millimeter for two extra
dimensions, without contradicting the laboratory data then available.
At distances shorter than the size of the extra dimensions, Newton's inverse-square
law would change, and the graviton would appear as a tower of
Kaluza--Klein modes.
Figure~\ref{fig:led}, adapted from the \emph{Scientific
American} article we wrote in 2000 \cite{sciam}, gives the picture.

So in 1998, just as the LEP paradox was taking hold, we had a radically different answer to the hierarchy problem: there was no hierarchy between the weak scale and the fundamental quantum-gravity scale. Gravity is weak because it spreads into the large extra-dimensional space, while the other forces do not, diluting its strength in the process. The price was a large compactification radius relative to the fundamental length---a new geometrical hierarchy whose origin remained unexplained, though it was radiatively stable \cite{admr}. This geometrical approach to the hierarchy problem changed where people looked for new physics. All of a sudden BSM physics no longer belonged mostly to high-energy physics and giant colliders. Tabletop searches for new physics got an enormous boost, especially tests of gravity at sub-mm distances. All the years of developing extraordinary new techniques at the precision frontier started paying large dividends. It helped a lot that the basic Large Extra Dimensions framework could be understood almost at a glance \cite{sciam}.

\FloatBarrier

\section{Anthropic reasoning in the multiverse}
\label{sec:anthropic}

A parallel story was unfolding in cosmology, one that would eventually change
the status of naturalness itself.

In 1987 Steven Weinberg wrote a paper \cite{weinberg87} that was
very controversial at that time.  
I happened to be visiting him on Black Monday, October 19, 1987, when the stock market crashed, and he told me about the paper. His argument was conditional on an ensemble of universes---what would later be called a multiverse---each with a different effective cosmological constant. Observers could arise only in those where the cosmological constant was small enough for structure to form.
If a positive vacuum energy density becomes sufficiently large relative
to the matter density when galaxies form, accelerated expansion prevents
gravitational collapse. Without galaxies there are no stars, no heavy
elements, and no observers to ask the question.  This was not a dynamical derivation, but an observer-selection
bound in an assumed multiverse. It could just as well have been called the Galactic Principle.

I confess that I left that meeting thinking that the great man had lost
his marbles.  He was fifty-four years old, which at the time seemed to me
quite advanced.  I have since revised my estimate of what counts as old, and my estimate of who was confused in that conversation.

A decade later, observations of distant Type Ia supernovae revealed
that the expansion of the universe is accelerating \cite{sn}.  The
inferred value was well below Weinberg's original upper bound, but it
lay in a range assigned appreciable probability by the more quantitative
anthropic calculation of Martel, Shapiro, and Weinberg, completed shortly
before the discovery \cite{weinberg87}.
To me, this was a striking success of anthropic reasoning. It did not prove the multiverse, and the inference remains conditional on what is assumed to scan and how the vacua are weighted. Even so, it made anthropic reasoning much harder to dismiss.

Meanwhile, string theorists were coming to terms with a related
surprise.  For years, one of the great hopes of string theory,
repeated in countless colloquia, was that the theory would be extraordinarily
predictive. Because it has no continuous dimensionless input parameters, it was tempting to believe that the electron-to-muon mass ratio, along with all the other dimensionless numbers of the Standard Model, would eventually be calculable. The lesson that emerged more slowly was simple in retrospect: a unique set of equations can still have many solutions.

Indeed, far from selecting a unique vacuum, string theory appears to
possess an enormous landscape of metastable vacua \cite{landscape}.  Each
vacuum describes a possible universe, with its own low-energy laws of
nature.  The cosmological-constant story and the landscape story had
converged. The multiverse became, at least logically, a plausible
structural feature of the most ambitious physical framework we had, and it provided a natural arena for anthropic reasoning.

This raised a new question: could anthropic selection apply not only to the cosmological constant, but also to the weak scale?

\section{Split Supersymmetry}

In January 1998, a very original paper by Agrawal, Barr, Donoghue, and Seckel \cite{agrawal} presented an anthropic argument for the weak scale based on the necessity of having both light and heavy atomic nuclei---later dubbed the Atomic Principle. I was captivated by it and showed it to Nima on our flight to Paris, but we were too preoccupied with extra dimensions to devote time to it then.

Fast forward to the early 2000s: Stanford was buzzing with activity on the multiverse \cite{landscape}.
I too was excited. Weinberg's argument and the 1998 discovery had already convinced me of the value of anthropic reasoning, and the multiverse provided a natural arena for it.
I felt then, and still do, that the multiverse is the most important consequence of string theory. So I began thinking about whether anthropic selection could be applied to the hierarchy problem as well. I was not content with the Agrawal et al.\ \cite{agrawal} proposal alone.
I could not believe that gauge coupling unification in supersymmetry was an accident. I wanted a framework that preserved gauge coupling unification while using the Atomic Principle to fix the weak scale.

A crucial inspiration came from the groundbreaking paper of Bousso and Polchinski \cite{landscape}. 
They showed that a large negative bare cosmological constant could be offset by positive, quantized contributions from many independent four-form fluxes. With sufficiently many fluxes, the allowed values form a dense discretuum near zero; in that high-dimensional flux lattice, increasing the magnitude of the bare term can make a near-cancellation easier.
In plain language, many independent, quantized contributions could make
an otherwise fantastically precise cancellation possible.
For me, this was a eureka moment: perhaps a large scale of SUSY breaking could coexist with a small weak scale after all. 
But even granting this, what about unification? How can one simultaneously have heavy superparticles and supersymmetric gauge coupling unification?

This was something that I had understood since 1981: 
Supersymmetric unification is largely insensitive to the masses of the scalar superpartners, squarks and sleptons, because they belong to complete grand unified theory (GUT) multiplets. By contrast, it depends predominantly on keeping the fermionic superpartners, gauginos and Higgsinos, light.
These come in incomplete GUT multiplets and shift the relative running of the three gauge couplings away from the Standard Model trajectory and toward successful GUT unification. Crucially, gauginos and Higgsinos can remain light when protected by appropriate approximate chiral symmetries, so their masses need not track the scalar supersymmetry-breaking scale.
This suggested a framework in which the superpartners are split: scalars are heavy at the SUSY-breaking scale, while fermions remain light near the weak scale. This also eliminates many of the phenomenological problems of the MSSM, such as flavor-changing neutral currents, which are tied to light squarks.

Armed with these insights, in the spring of 2004 I went for a long-planned extended visit to Harvard where Nima had been for some time. When I arrived, the atmosphere was decidedly hostile to the multiverse. Nima was vehemently opposed to it and wanted to entice me to work on ghost condensation, an original idea that he hoped might address the cosmological constant problem. After a week of vigorous arguing, he gradually warmed to the multiverse, and then we had an extraordinary time working nonstop on it.

As we worked, we came across a closely related paper from the previous year by James Wells \cite{Wells:2003tf}
 in which he proposed a ``Little Hierarchy'' between gauginos and scalars as a way to ameliorate the phenomenological problems of the MSSM and account for the LEP paradox. Soon after our paper appeared, Gian Giudice and Andrea Romanino published a detailed study of the framework \cite{split}, which they tastefully named Split Supersymmetry, Split for short.

One of our primary objectives was to identify distinctive signatures of Split. For sufficiently high scalar masses, a particularly striking one is a long-lived gluino. Its ordinary decays proceed through virtual squarks and are strongly suppressed; after hadronizing into an R-hadron, it can produce displaced vertices or slow, charge-changing tracks in a detector \cite{split}. These signatures are targets for current searches and will remain so at the high-luminosity LHC\@. There are a number of other predictions of Split that I will not discuss here.

In short, Split preserves several of the great virtues of
supersymmetry---gauge coupling unification, dark matter candidates, and a
connection to very high-scale physics---while removing many of the
difficulties that had made the MSSM uncomfortable.  The ease with which
this works is striking.  It does not prove that Split is right, of
course, but it suggests that the idea may contain a germ of truth.  If I
were forced today to name the explanation I find most likely for what
Nature is doing, I would say Split Supersymmetry.

There are obvious reasons why Split is not more popular.  One is its
possible connection to the multiverse.  Another is that the colliders
with the best chance of testing it may be half a century away.  That
timescale has a paralyzing effect, not only on Split, but on any BSM
theory whose decisive predictions lie well beyond the TeV scale.  Imagine
a world in which a 100~TeV collider were scheduled to turn on this
December.  Interest in Split would surge, as it would for other
multi-TeV models, and detectors and collider parameters would be feverishly
optimized with such possibilities in mind.

Split Supersymmetry also has something else: a possible underlying philosophy---the same anthropic philosophy that, in my view, offers our most compelling framework for the cosmological constant problem.

\section{A quarter century later}

We have lived through two cycles of high hopes, suggestive evidence,
and contraction: grand unification in the 1980s, and supersymmetric optimism followed by the LEP
paradox in the 1990s and beyond.  We have seen the cosmological
constant evolve from an intractable embarrassment into a possible
window onto the multiverse.  We have seen the principle of naturalness,
which organized the field from 1978 onward, called into question even
by some of the people who built the field around it.

In the quarter century since the period covered by this symposium,
the most important discovery in particle physics was that of the Higgs boson in 2012. At colliders, the Standard Model has triumphed, while the simplest natural weak-scale solutions to the hierarchy problem remain unconfirmed and severely constrained. The LEP paradox became the LHC paradox. Naturalness, as applied to the weak scale, has lost still more of its authority. Perhaps in retrospect this should not have been too surprising, given the far more drastic tuning required for the cosmological constant, for which no generally accepted dynamical solution has been found.

To me, the success of anthropic reasoning for the cosmological constant, together with the emergence of the string-theory multiverse, makes a compelling case for exploring anthropic approaches to the hierarchy problem.

Many physicists, even string theorists, view the multiverse with suspicion or worse.
This recalls a view common a few centuries ago: that there was only one solar system. The analogy is imperfect: other planetary systems eventually became directly observable, whereas empirical access to a multiverse is far more indirect. Still, experiment and observation are the ultimate arbiters, and they can take time. The recognition of a multitude of galaxies, and with them $\sim 10^{22}$ solar systems, took a few centuries and enormous technological progress built on advances across physics.

\textit{How else might we test multiverse reasoning?} Discovery of Split SUSY would strengthen the case for environmental selection. A further possibility comes from axions: the topological complexity of string compactifications can produce a multitude of light axions.

These ``Axiverse'' axions are being sought in laboratory experiments and through their effects on rotating black holes, including gravitational waves from black-hole superradiance \cite{axiverse}. Discovering several would be striking evidence for an axiverse and would support the topological richness expected in string compactifications. Neither discovery would by itself establish a multiverse.

I do not know whether the next turn of the curve in
Figure~\ref{fig:rollercoaster} is upward or downward.  The terminal
question mark is honestly drawn.  Future colliders, precision Higgs
measurements, flavor experiments, a host of tabletop
experiments---including electric-dipole-moment searches, dark-matter experiments,
and gravitational tests at short distances---as well as gravitational-wave
and cosmological observations may each supply part of the answer.  We may
find supersymmetry after all, perhaps in a form less natural than the
one we first hoped for.  We may find a deviation in Higgs couplings, a
long-lived particle, a dark-matter signal, or an unexpected failure of
gravity at short distances.  Or we may be forced to accept that the
hierarchy problem is telling us something more radical than any of the
solutions we have so far proposed.

What I do believe is that the questions are durable.  Why is gravity so
weak?  Why is the vacuum energy so small?  Why are there three families
of quarks and leptons, with the masses and mixings they have?  These
questions have outlived many proposed answers.  That should not depress
us.  It is what gives the field its peculiar charm.  The history of BSM
physics is not only a history of theories that have not yet been
confirmed.  It is also a history of questions so sharp that they keep
forcing us to enlarge our idea of what an answer could be.

\section*{Acknowledgements}

For all the fun we had doing physics together, I thank my collaborators in the work described here: Ignatios Antoniadis, Nima Arkani-Hamed, Asimina Arvanitaki, Sergei Dubovsky, Gia Dvali, John Ellis, Howard Georgi, Gian Giudice, Nemanja Kaloper, John March-Russell, Stuart Raby, Leonard Susskind, and Frank Wilczek. I thank Peter Michelson for pointing me to the untested centimeter.

I thank the organizers for a superb history symposium and hope the next,
covering particle physics in this millennium, will happen soon. This work
was supported in part by NSF Grant PHY-2310429 and the
Leinweber Institute for Theoretical Physics at Stanford.

\end{document}